# OLEDs as models for bird magnetoception: detecting electron spin resonance in geomagnetic fields

Tobias Grünbaum,[1] Sebastian Milster,[1] Hermann Kraus,[1] Wolfram Ratzke,[1] Simon Kurrmann,[1] Viola Zeller,[1] Sebastian Bange,[1] Christoph Boehme,[2] and John M. Lupton[1]

[1]*Institut für Experimentelle und Angewandte Physik, Universität Regensburg, Universitätsstraße 31, 93053 Regensburg, Germany*

[2]*Department of Physics and Astronomy, University of Utah, 115 S, 1400 E, Salt Lake City, UT 84112, USA*

Certain species of living creatures are known to orientate themselves in the geomagnetic field. Given the small magnitude of approximately 48 µT, the underlying quantum mechanical phenomena are expected to exhibit coherence times approaching the millisecond regime. In this contribution, we show sensitivity of organic light-emitting diodes (OLEDs) to magnetic fields far below Earth's magnetic field, suggesting that coherence times of the spins of charge-carrier pairs in these devices can be similarly long. By electron paramagnetic resonance (EPR) experiments, a lower bound for the coherence time can be assessed directly. Moreover, this technique offers the possibility to determine the distribution of hyperfine fields within the organic semiconductor layer. We extend this technique to a material system exhibiting both fluorescence and phosphorescence, demonstrating stable anticorrelation between optically detected magnetic resonance (ODMR) spectra in the singlet (fluorescence) and triplet (phosphorescence) channel. The experiments demonstrate the extreme sensitivity of OLEDs to both static as well as dynamic magnetic fields and suggest that coherent spin precession processes of Coulombically bound electron spin pairs may play a crucial role in the magnetoreceptive ability of living creatures.

## Introduction

Organic light-emitting diodes (OLEDs) constitute unique material and device systems which allow for the investigation of the spin permutation symmetry of electron-hole pairs as well as their coupling to nuclear spins at ambient conditions [1-4]. Even magnetic fields corresponding to energy scales many orders of magnitude below the thermal energy have been shown to alter the dynamic equilibrium between singlet and triplet charge-carrier pairs.

Such exceptional sensitivity to local magnetic fields is similar to the orientational ability of certain migratory bird species, which are able to detect miniscule changes in Earth's magnetic field [5]. The common ground of these two seemingly very different organic systems is that radical-pair processes are involved to explain the magnetic sensitivity of both OLEDs and many species of living creatures.

Apart from several models based on the incorporation of magnetite in animal bodies [6, 7], a number of approaches propose that magnetoception arises as a consequence of magnetically sensitive chemical reactions [8-12]. In this model, the reaction products depend on the balance between singlet-like and triplet-like charge-carrier pairs as a consequence of spin mixing, which can be influenced by both static as well as dynamic magnetic fields. This interpretation is supported by studies which showed that birds become disorientated by exposure to RF irradiation with frequencies in the few-MHz regime [10, 13, 14]. Both experimental as well as theoretical studies suggest electron spin coherence times exceeding 100 µs to explain these findings [15, 16]. Please note that different notions of "spin coherence", i.e. the sustained relative orientation of the spins of an individual charge-carrier pair, are employed in literature [15-17]. In this contribution, we refer to the timescale of a fixed phase relation of the individual charge-carrier spins comprising a given carrier pair as the "coherence time". We emphasize, though, that this definition does not imply a fixed orientation of electron spins between different charge-carrier pairs.

The strong magnetic field sensitivity of OLEDs indicates spin coherence times in a similar regime, since the spin coherence time determines the magnitude of changes in magnetic field which can be reflected in a DC observable such as resistance or electroluminescence [17, 18]. The precession of two spins can only be distinguished if their individual phases are preserved for at least the inverse of the difference of their Larmor precession frequencies. As the Larmor frequency is proportional to the magnetic field, the sensitivity of an OLED to a given magnetic-field step allows a lower bound for the spin coherence time to be deduced. Using a solid-state OLED device comes with the advantage that non-equilibrium, coherent phenomena can be directly probed in the current by means of electron paramagnetic resonance techniques [1, 19, 20]. Moreover, this approach can be applied to detect resonant changes in the singlet-triplet equilibrium in a time-averaged manner with various observables available for detection such as device resistance and electroluminescence, the dielectric constant of the emitting layer, or chemical reaction yields [21-27].

## Results and Discussion

**Static magnetic field effects in OLEDs**

As in the case of birds sensing magnetic fields in the µT range, we show the sensitivity of an OLED to magnetic fields far below Earth's magnetic field. OLED devices were fabricated on glass substrates coated with a 100 nm thick indium tin oxide (ITO) layer from which a stripe-shaped anode was defined by etching. A spin-coated 80 nm thick layer of poly(styrene-sulphonate)-doped poly(3,4-ethylene-dioxythiophene) (PEDOT:PSS) serves as a hole-injection layer. In a next step, 90 nm of the commercial poly(phenylene-vinylene) (PPV) copolymer "SuperyellowPPV" (SyPPV, Merck AG) was spin-coated, before a 3 nm thick barium cathode and a 250 nm aluminium capping layer were deposited by thermal evaporation. The devices were then encapsulated inside a nitrogen glovebox with two-component epoxy glue and a glass cover slip to prevent oxidation of the polymer and metal layers. The typical turn-on voltage of a SyPPV OLED at room temperature is approximately 2.5 V with bright yellow fluorescence peaking around an emission wavelength of 540 nm. For the experiments discussed in the following, if not stated otherwise, we operate the SyPPV device at a constant current of 100 µA, corresponding to a current density of 3.5 mA/cm$^2$. All experiments were performed at room temperature.

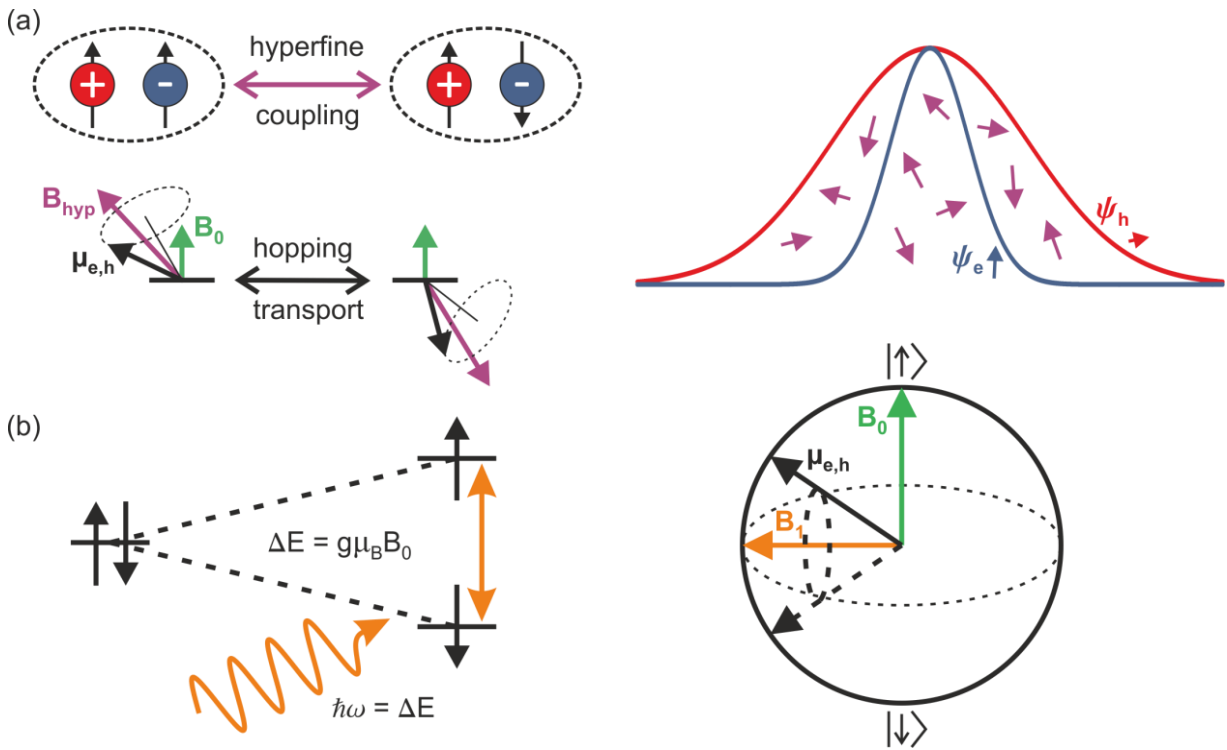

*Fig. 1: Hyperfine coupling within the radical-pair model of OLEDs (a). At zero or low external magnetic field $B_0$, the orientation of the magnetic moment µ of a charge carrier – electron (e) or hole (h) – is predominantly determined by the coupling to magnetic moments of surrounding hydrogen nuclei. This coupling is expressed as an effective hyperfine field $B_{hyp}$. Electrons and holes experience different hyperfine fields due to the differing localization of their respective wavefunctions. A more delocalized wavefunction leads to averaging over a larger amount of randomly oriented magnetic moments and therefore to a smaller effective $B_{hyp}$. The red and the blue arrow indicate the resulting hyperfine field for the broader and the narrower wavefunction. (b) Application of a static external magnetic field leads to a stabilization of the spin configuration and to a Zeeman splitting ΔE of spin-up and spin-down states of an individual charge carrier. RF excitation resonant with ΔE induces transitions between the Zeeman-split states. On a Bloch sphere and in the co-rotating frame, this effect can be visualized as the precession of the magnetic moment of the charge carrier around the excitation field $B_1$.*

The inherent magnetic-field sensitivity of bipolar OLEDs is explained by the radical-pair model [3, 21, 28] of magnetoresistance (MR) and magnetoelectroluminescence (MEL) illustrated in Fig. 1. Following the injection of electrons and holes from the electrodes, the charge carriers are transported through the emitting polymer layer via hopping processes. If two charge carriers are located on neighbouring polymer chains, a Coulombically bound electron-hole pair can be formed. Depending on the relative spin orientation of its constituents this pair can exist in either two product states configurations with each approximately half singlet and half triplet content, or two product state configurations with pure triplet-like states, causing an overall singlet to triplet distribution of 1:3 according to simple spin statistics, as thermal polarization can be neglected entirely under low magnetic field conditions. Further localization of both charge carriers on a single polymer chain leads to the creation of either pure singlet or pure triplet excitons characterised by a strong exchange interaction. While singlet excitons decay to the singlet ground state radiatively on a nanosecond timescale, a process known as fluorescence, emission from the triplet state, i.e. phosphorescence, is dipole forbidden in pure singlet emitters such as SyPPV.

At vanishing external magnetic fields, the spin orientation of a charge carrier is predominantly determined by the effective hyperfine field originating from the randomly oriented magnetic moments of the abundant hydrogen nuclei in the emitting polymer layer [29]. As a consequence, the spin configuration of a charge-carrier pair can change from parallel (singlet) to antiparallel (triplet) due to the different hyperfine field directions it is subjected to during hopping transport as indicated in Fig. 1a. Application of an external magnetic field $B_0$ exceeding the hyperfine fields stabilizes the spin configuration of a given charge carrier, thus leading to a suppression of the hyperfine field-induced intermixing between singlet-like and triplet-like states. As the rates for dissociation of electron-hole pairs to free charge carriers and for recombination to excitons differ for singlet-like and triplet-like configurations of the pairs [26, 28, 30], a change of both device resistance and electroluminescence (EL) arises upon application of an external magnetic field. MR and MEL of OLEDs have been studied in detail at both cryogenic and ambient temperatures at various magnetic field regimes [4, 28, 31-40]. Under magnetic resonance, illustrated in panel 1b, spin-flips of individual charge carriers are induced, giving rise to a mixing of singlet and triplet pair configurations.

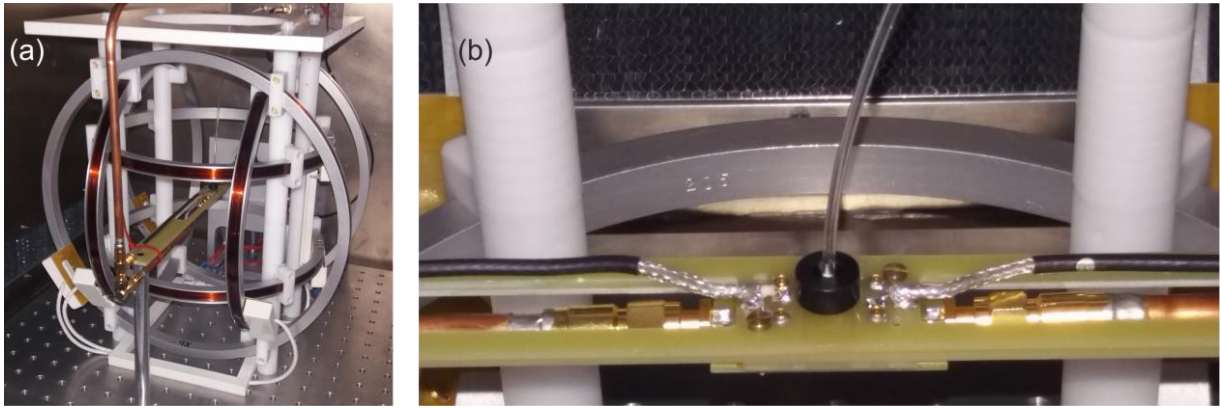

*Fig. 2: Experimental setup. In the low-field regime (< 2 mT), the external magnetic field is applied by a set of Helmholtz coils (a). This procedure also allows for the compensation of Earth's field. The sample holder is shown in (b). It has connections for RF excitation via a coplanar stripline and the DC current supply of the OLED. An optical fibre is attached from the top to collect electroluminescence.*

For the experiments discussed in this paper, the external magnetic field $B_0$ is provided either by a 1D set of Helmholtz coils for fields up to 30 mT or by a 3D array of Helmholtz coils (Ferronato BH300-3-A, Fig. 2a) for fields up to 2 mT, which can be applied in an arbitrary direction. A photograph of the latter setup is shown in Fig. 2. With the 3D Helmholtz coils, Earth's magnetic field is compensated in the measurements presented by applying a field of equal amplitude but antiparallel orientation. For each coil, a separate bipolar power supply

(CAEN ELS easy driver 5020) was used. A Keithley 238 source measure unit operates the SyPPV OLED under constant current conditions and simultaneously measures the DC device resistance. Luminescence is collected with an optical fibre and detected with a Femto OE-200-SI optical power meter, which in turn is read out by a Keysight 34461A multimeter. The sample holder including the optical fibre and the leads for resistance measurement is shown in Fig. 2b.

Application of a magnetic field leads to a systematic decrease of device resistance as well as to a simultaneous increase of singlet luminescence (fluorescence) on the scale of a few percent. An example of this effect is shown in Fig. 3a. At fields below 1 mT (see Fig. 3b), the shape of the MR and MEL curves is inverted as a consequence of the ultra-small magnetic field effect originating from dipolar coupling of charge-carrier spins to other electronic and nuclear spins [37, 40]. Coupling of electronic and nuclear spins leads to the emergence of partially degenerate sublevels, which shift relative to each other upon application of a magnetic field. The inversion of the curve shapes is thought to result from an increased mixing of singlet-like and triplet-like charge-carrier pairs due to crossings of the hyperfine-split spin sublevels at non-zero $B_0$ [37]. For fields of order $B_{Earth}$, we are able to resolve the MR of an OLED on the scale of 0.1 parts per million (ppm), yielding a magnetic-field sensitivity of 300 nT at room temperature as shown in Fig. 3c. To observe the effect of a magnetic field on a direct current (DC) observable, e.g. device resistance or EL, the spin configuration of a charge-carrier pair must be measured for at least the inverse of the difference in Larmor frequencies. A magnetic-field change of 300 nT corresponds to a difference in Larmor frequencies of approximately 10 kHz. Consequently, the observed sensitivity implies coherence times approaching 100 µs for the spins of charge-carrier pairs in an OLED under ambient conditions, a value similar to the coherence times invoked to explain the magnetoreceptive abilities of certain bird species [15-17].

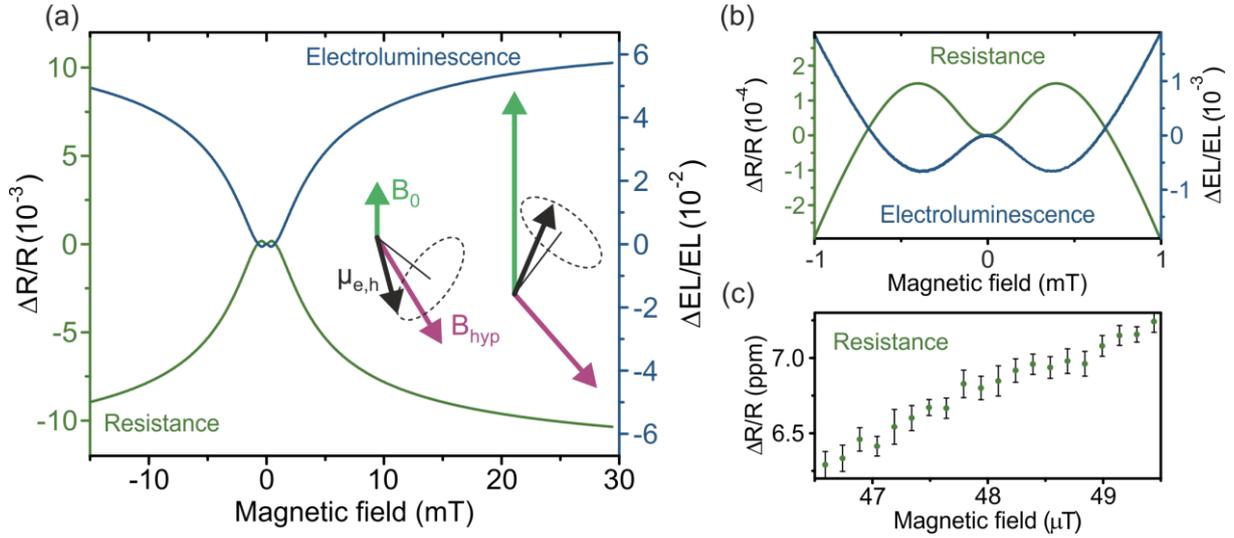

*Fig. 3: Room-temperature magnetic-field sensitivity of the SyPPV OLED. (a) Under constant current conditions at I = 100 µA, the device resistance shows a decrease upon the application of an external magnetic field on the mT scale, while the electroluminescence increases. The arrows schematically indicate the relative orientations of the external field $B_0$, the electronic magnetic moments $\mu_{e,h}$ and the effective hyperfine fields $B_{hyp}$. In (b), the same data are plotted for $B_0$ from -1 to 1 mT, revealing the ultra-small magnetic field effect. (c) A zoom into the field region of geomagnetic field strengths reveals sub-microtesla magnetic-field sensitivity of the device with resistance changes in the 0.1 parts per million (ppm) regime resolved. The vertical error bars denote the standard deviation of the mean obtained from averaging of 30 data points with a measurement time of 1 s each. We note that, given the uncertainty in positioning the Hall probe used as well as its limited sensitivity, there is a finite uncertainty in determining the magnetic field. This error, which is of order 1 µT, does not change with magnetic field strength and is therefore not indicated in the figure. The uncertainty of the current-to-magnetic-field conversion factor of the Helmholtz coils is smaller than the size of the data points.*

**Electron paramagnetic resonance spectroscopy of a SyPPV OLED**

An experimental approach explicitly relying on the coherence of electron-hole pairs is electron paramagnetic resonance (EPR). Here, radio frequency (RF) excitation drives resonant transitions between the Zeeman-split energy levels of an individual charge-carrier spin as illustrated in Fig. 1b. Thereby, within the radical-pair picture, singlet-like charge-carrier pairs are transferred to the triplet manifold and *vice versa*. Following the same arguments as detailed above for the DC magnetoresistance and MEL, this change of singlet-triplet population ratio allows the detection of EPR in the device resistance (i.e. electrically detected magnetic resonance, EDMR) as well as in electroluminescence (optically detected magnetic resonance, ODMR). We perform EPR experiments by exciting a SyPPV OLED sample with a coplanar stripline designed to match a 50 Ω impedance. The RF signal provided by an Anritsu MG3740A signal generator is amplified to power levels up to 20 W by a HUBERT A 1020 RF amplifier and fed through the coplanar stripline producing the resonant excitation field $B_1$. The RF excitation transmitted through the stripline is routed out of the measurement setup into an impedance-matched line termination. The static magnetic field $B_0$ is applied such that $B_0 \perp B_1$.

Measurements of device resistance and luminescence under RF excitation at an excitation frequency of $f = 280$ MHz and a power of $P = 8$ W reveal distinct peaks around the resonance field $B_{EPR} = \pm 10$ mT as shown in Fig. 4a. These EDMR and ODMR peaks counteract the static MR and MEL behaviour, as resonant RF excitation re-opens a mixing channel between the singlet and the triplet manifold which was suppressed by the application of $B_0$ – thereby influencing both the density of free charge carriers as well as that of luminescent excitons. The overall magnitude of the resonance peaks in continuous-wave EDMR and ODMR arises from the time integral of the spin-pair dynamics under resonance, which typically occurs on the microsecond timescale [19, 30]. In EDMR, magnetic resonance transforms long-lived triplet pairs into short-lived singlet pairs, initially enhancing the current. But some singlet pairs are also converted into triplets, so on long timescales, a current quenching is observed in time-resolved experiments under pulsed resonant excitation. The steady-state resonance features are therefore a superposition of quenching and enhancement features, which complicates a quantitative analysis of the peak amplitude. In Fig. 4b, curves with and without RF excitation are compared. Apart from the resonant signal, a minor non-resonant contribution to both resistance as well as EL intensity is observed. This deviation under non-resonant drive originates from the RF-induced temperature increase of the OLED. Because of the thermally assisted hopping transport in the OLED, the resistance decreases with increasing temperature, resulting in a slight change of magnetic-field effects off resonance.

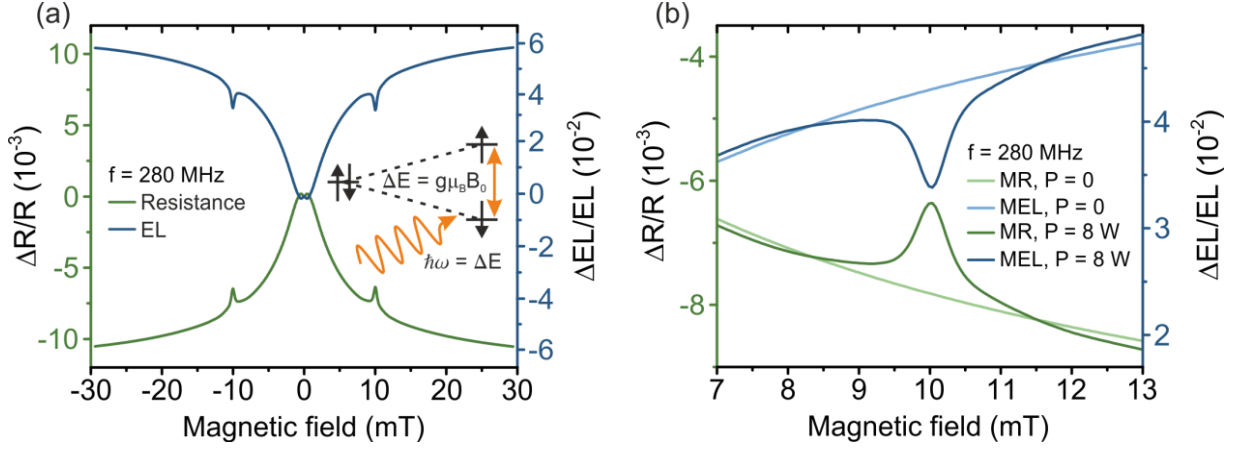

*Fig. 4: (a) Magnetoresistance (green) and magnetoelectroluminescence (MEL) (blue) of the SyPPV OLED under RF excitation at f = 280 MHz and P = 8 W. Resonant peaks at $B_{EPR}$ = 10 mT are observed in both resistance and electroluminescence. The sketch illustrates the Zeeman splitting of the spin-½ species. (b) The resonant peaks are superimposed on the MR and MEL curves from Fig. 3a, revealing a slight non-resonant contribution due to the RF-induced temperature increase of the sample.*

To increase the signal-to-noise ratio, lock-in detection with 100% square wave modulation of the RF excitation power at a frequency of $f_{mod}$ = 232 Hz is employed. With this technique, the EDMR and ODMR signal is observed without the static MR or MEL background as seen in Fig. 5a, which allows for reliable lineshape analysis of the EPR signal. At sufficiently moderate excitation power, the resonance lines are accurately described by the sum of two Gaussian functions with identical centre positions but different widths – as indicated in Fig. 5b –, corresponding to the different distributions of hyperfine fields experienced by electron and hole and the resulting inhomogeneous broadening of the resonance transition [25, 41]. While the data presented here do not allow the assignment of the individual Gaussian peaks to a certain charge-carrier species directly, this becomes possible in high-field EDMR experiments where the different g-factors of electron and hole can be resolved [39]. Fitting the model of two Gaussians to the experimental data yields linewidths (FWHM) of 0.44 mT for the narrow hole peak and 2.31 mT for the broader electron spectrum. A sample EDMR spectrum taken at $P$ = 20 mW with the corresponding fit is displayed in Fig. 5b. The EPR resonance condition is given by $\hbar\omega = \Delta E = g\mu_B B_{EPR}$ with the excitation frequency omega, the g-factor $g$, the Bohr magneton $\mu_B$ and the resonance field $B_{EPR}$. We performed EPR measurements for frequencies between 0.1 and 400 MHz at an excitation power of 8 W, corresponding to an approximate magnetic-field amplitude of the RF excitation wave of $B_1$ =

0.1 mT. By fitting EDMR as well as ODMR spectra taken at a range of excitation frequencies, the linear relationship between $\omega$ and $B_{EPR}$ is verified in the inset of Fig. 5a. The data are accurately described by the free-electron g-factor of $g = 2.0023$.

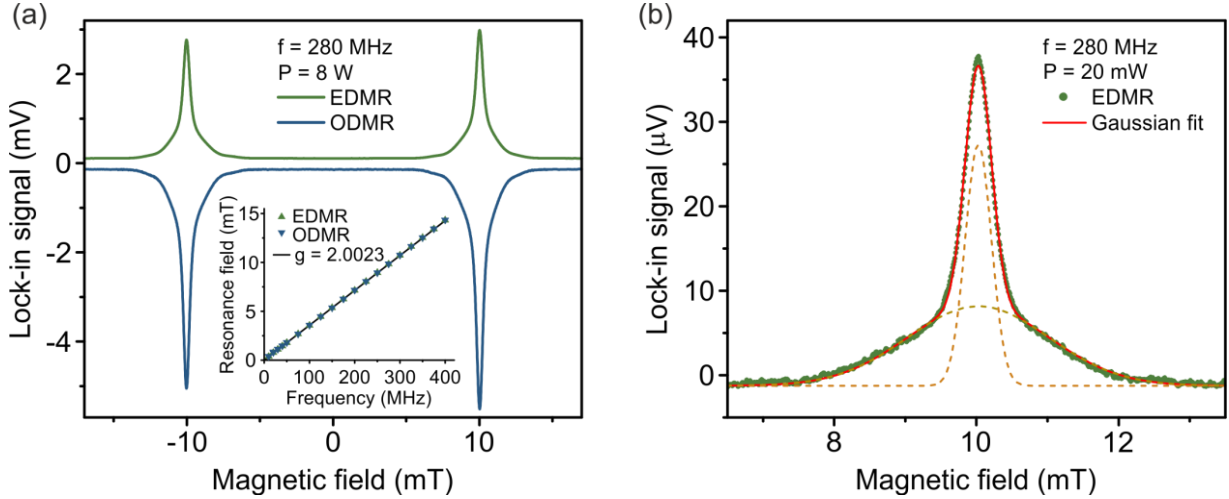

*Fig. 5: Electrically detected magnetic resonance (EDMR) and optically detected magnetic resonance (ODMR) of the SyPPV OLED measured by lock-in detection. (a) Sample EDMR (green) and ODMR (blue) taken at f = 280 MHz and P = 8 W featuring a resonance field $B_{EPR} = 10$ mT. The inset shows the frequency dependence of $B_{EPR}$ for values between 10 and 400 MHz. EDMR and ODMR resonance fields are indicated by green and blue triangles and show excellent agreement. The data are accurately described by the EPR resonance condition with the free-electron g-factor g = 2.0023 (black line). (b) EDMR spectrum taken at P = 20 mW (green) fitted by the sum of two Gaussian functions (red solid line), representing the inhomogeneous broadening of the electron and hole resonance by hyperfine fields. Excellent agreement with the experimental lineshape is found. The individual Gaussian functions are depicted as dashed lines.*

In addition, the power dependence of the EDMR signal can be investigated by performing EPR measurements at an excitation frequency of $f = 280$ MHz for excitation powers ranging from 2 mW up to 20 W. At elevated RF power levels, a significant broadening of the EDMR spectra is observed in Fig. 6a. The reason for this power broadening is the saturation of the Lorentzian absorption spectrum of a two-level system, which remains in the excited state for longer times with increasing driving field $B_1$ [42]. This saturation becomes apparent when the Lorentzian width of the resonant transition becomes comparable or even overcomes the Gaussian inhomogeneous broadening of the hyperfine field strengths. As a consequence, for high excitation power, resonance spectra are no longer given by the sum of two Gaussians representing the hyperfine distributions of electron and hole. Instead, the lineshape is described by the sum of the two Gaussian peaks convoluted with a Lorentzian, which is known as the Voigt profile. Application of this model to measurements at different power levels (not shown) provides an estimate of the excitation field strength $B_1$ as detailed in [43]. At an excitation power of 20 W, we find a driving field $B_1$ of approximately 0.2 mT, which is in good agreement with calculations based on the geometry of the coplanar stripline used for RF excitation.

Upon increasing the RF excitation power, the amplitude of the EDMR and ODMR peaks increases linearly until a transition to sublinear behaviour at powers surpassing 100 mW is observed in Fig. 6b. This transition is in agreement with theoretical predictions for electron-hole pairs under strong resonant RF driving made in Refs. [44, 45] and the corresponding observations reported in Refs. [43, 46] including strongly non-linear behaviour at high driving fields. In brief, the transition to a dependence of the amplitude on the power $P$ as $\sqrt{P}$ arises due to power broadening. Even lower exponents and an inversion of the sign of the resonance signal occur due to the onset of collective precession of the resonant species, the superradiant spin-Dicke effect [46]. We find that the degree of nonlinearity for a given power level is stronger for the narrower hole Gaussian, i.e. for charge carriers experiencing weaker hyperfine fields on average due to their delocalization over many hydrogen nuclei leading to cancellation effects [28]. This observation is to be expected since the onset of the spin-Dicke effect occurs at lower driving fields $B_1$ in deuterated materials with weaker hyperfine interactions than in conventional protonated compounds.

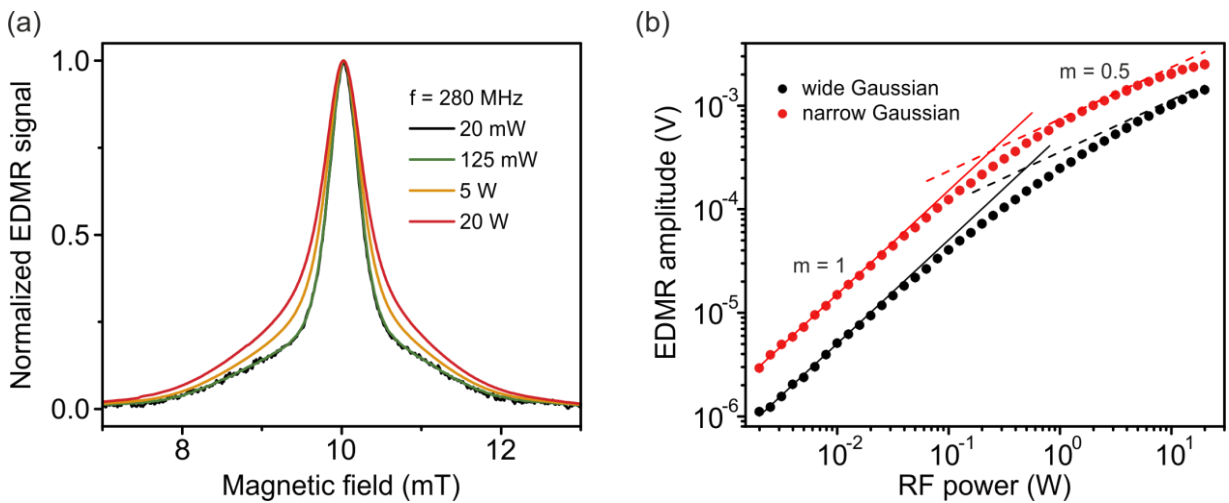

*Fig. 6: Power dependence of the lock-in detected EDMR signal of SyPPV OLEDs. In (a), spectra taken at f = 280 MHz and power levels of 20 mW, 125 mW, 5 W and 20 W are shown. While the linewidth is constant for weak excitation, an increase of linewidth at elevated powers due to power broadening is observed. (b) Power dependence of the EDMR amplitude. The amplitudes of the wide and narrow Gaussian functions as defined in Fig. 5(b) are depicted as black dots and red dots, respectively. A transition from a linear increase of the amplitude (indicated by solid lines) at low excitation power to strongly sublinear behaviour at high power is observed. The $\sqrt{P}$-dependent increase is indicated by dashed lines. ODMR spectra exhibit the same power-dependent behaviour (not shown).*

In order to obtain a lower bound for the coherence time of the radical pairs in an OLED, the excitation frequency under resonant drive is decreased to the scale of a few MHz. Remarkably, even under these conditions, summarized in Fig. 7, the Zeeman resonances are resolved as distinct peaks in both EDMR and ODMR (cf. Fig. 7a, b) down to a frequency of 7 MHz. This frequency corresponds to a resonance field $B_{EPR} = 0.25$ mT, which is only five times the value of the geomagnetic field. Observing a resonance at this frequency implies that the spin coherence time must equal at least 1/7 MHz = 140 ns at room temperature. This value constitutes an absolute lower bound to the coherence time, which is most likely much larger given the exquisite sensitivity of spin-dependent recombination to static magnetic field changes as discussed above. When decreasing the excitation frequency to even lower values, an additional peak centred around $B_0 = 0$ is observed in the EDMR and ODMR spectra, inhibiting a clear resolving of the Zeeman resonances, which are only visible as shoulders of the zero-field peak.

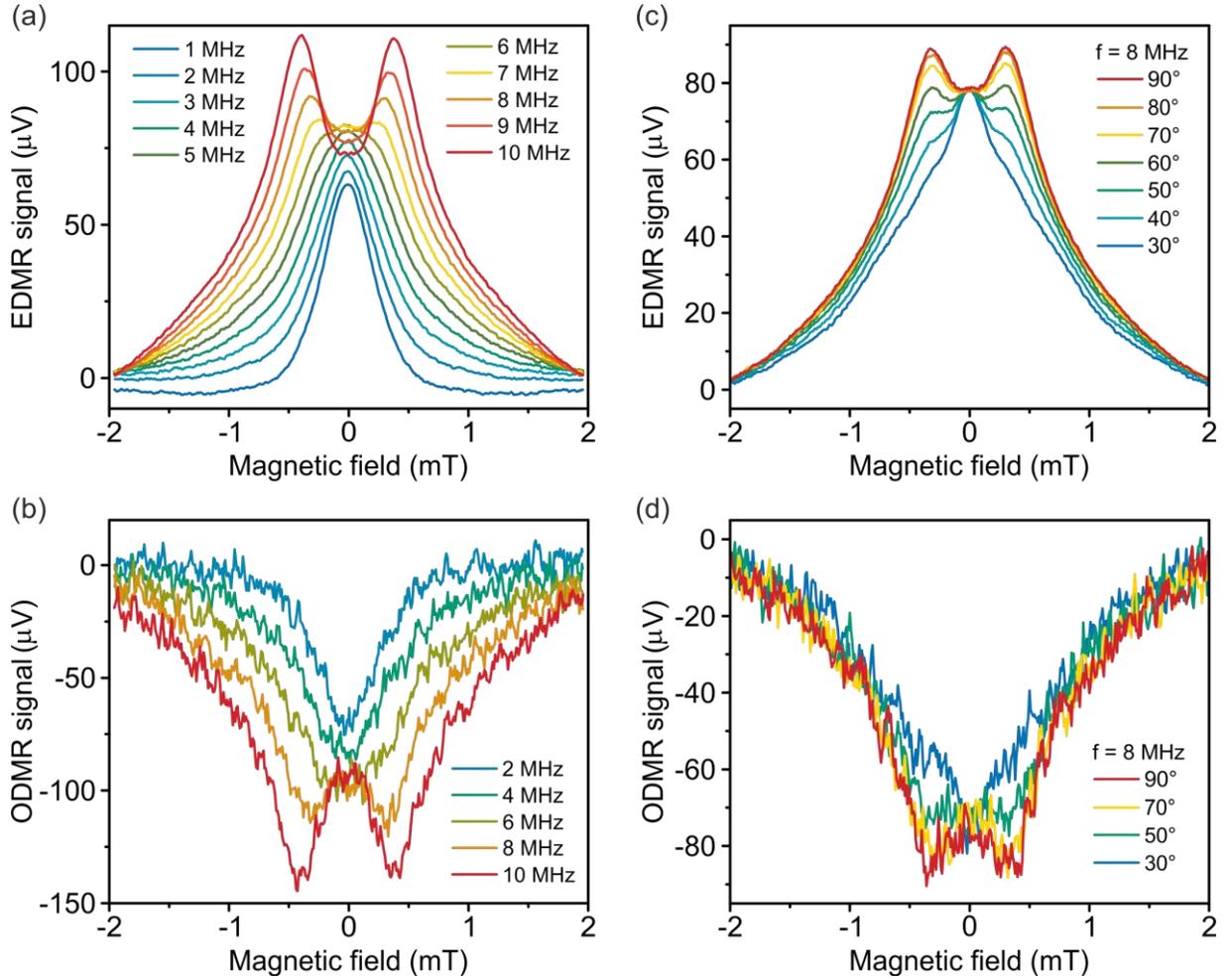

*Fig. 7: Low-frequency EDMR and ODMR spectra of SyPPV recorded by lock-in detection. (a) With decreasing excitation frequency f, the Zeeman resonances disappear and are overwhelmed by a new peak emerging at $B_0 = 0$. The amplitude of this zero-field peak is essentially constant below f = 4 MHz. By rotation of the excitation geometry (c), the hyperfine-based origin of the zero-field peak is verified. For f = 8 MHz, the Zeeman resonances are resolved symmetrically around 0 mT under perpendicular orientation of $B_0$ and $B_1$. Rotation of the external magnetic field towards a configuration parallel to $B_1$ leads to suppression of the g = 2 Zeeman resonances, rendering the zero-field peak clearly visible. The numbers indicate the angle between $B_0$ and $B_1$, which is controlled by the low-field vector magnet shown in Fig. 2. For ODMR spectra (b, d) the same behaviour is observed qualitatively. Here, only every second curve is displayed for increased visibility.*

**Low-frequency EDMR and ODMR spectroscopy**

We now turn our analysis to the zero-field peak, which is observed both in EDMR and ODMR at sufficiently low excitation frequencies. We propose that this peak arises due to resonant transitions between Zeeman-split levels induced by the locally varying hyperfine field even at vanishing external magnetic field $B_0$. In other words, at sufficiently low frequencies, spin pairs will always exist which are in resonance with the driving field because of the intrinsic Zeeman splitting arising from the hyperfine fields. This bears conceptual similarity to birds becoming disorientated upon exposure to RF irradiation with $f = 7$ MHz corresponding to a larger energy splitting than the Zeeman splitting induced by Earth's magnetic field [10, 14]. To test this assumption, we rotate the excitation geometry from $B_0 \perp B_1$ towards $B_0 \parallel B_1$. While the Zeeman resonances are expected to disappear for the latter case due to the vanishing transition dipole moment, a resonance peak originating from the isotropic local hyperfine fields should not be affected by orientation [24, 47]. In the experiment, we gradually rotate the field from perpendicular towards parallel excitation by turning the field-sweep axis with the 3D Helmholtz coils. The results displayed in Fig. 7c, d show that at a frequency of $f = 8$ MHz we indeed observe a complete suppression of the Zeeman resonances at $B_0 = 0.29$ mT in EDMR and ODMR, whereas the zero-field peak amplitude remains virtually constant for all angles.

For further characterization of the zero-field peak, we investigate both its dependence on driving power as well as on OLED current density. We find that the strength of the zero-field peak increases strongly upon increasing the RF excitation power, with a similar but weaker dependence found for increasing device current as displayed in Fig. 8. Both observations can be explained qualitatively by assuming that both increased excitation power as well as a higher device current give rise to a sampling of a larger subensemble of the hyperfine distribution. By choosing these parameters such as to minimize the amplitude of the zero-field peak, $g = 2$ Zeeman resonances can be observed even at a frequency as low as 5.5 MHz at $B_{EPR} = 0.2$ mT, as is shown in Fig. 8c. As the width of the resonance peaks is generally determined by the hyperfine field strengths [46], we anticipate that the use of deuterated PPV-based materials as OLED emitters will allow us to decrease the detection limit of the EPR peaks to even lower values of the excitation frequency.

With these observations providing support for the interpretation that the zero-field peak is based on hyperfine-mediated spin mixing, the question arises whether the hyperfine field distribution can be estimated from the width of this peak. To minimize any additional broadening by the influence of the Zeeman resonances, EDMR and ODMR measurements were performed at $f = 0.5$ MHz and $I = 100$ µA as shown in Fig. 8d, e. By fitting both peaks with a Gaussian function, we find the FWHM to be 0.49 mT. This value is in remarkably good agreement with the narrower Gaussian linewidth extracted from EDMR spectra in the 100 MHz regime in Fig. 5b, corroborating the assumption of a hyperfine-based origin of the zero-field peak.

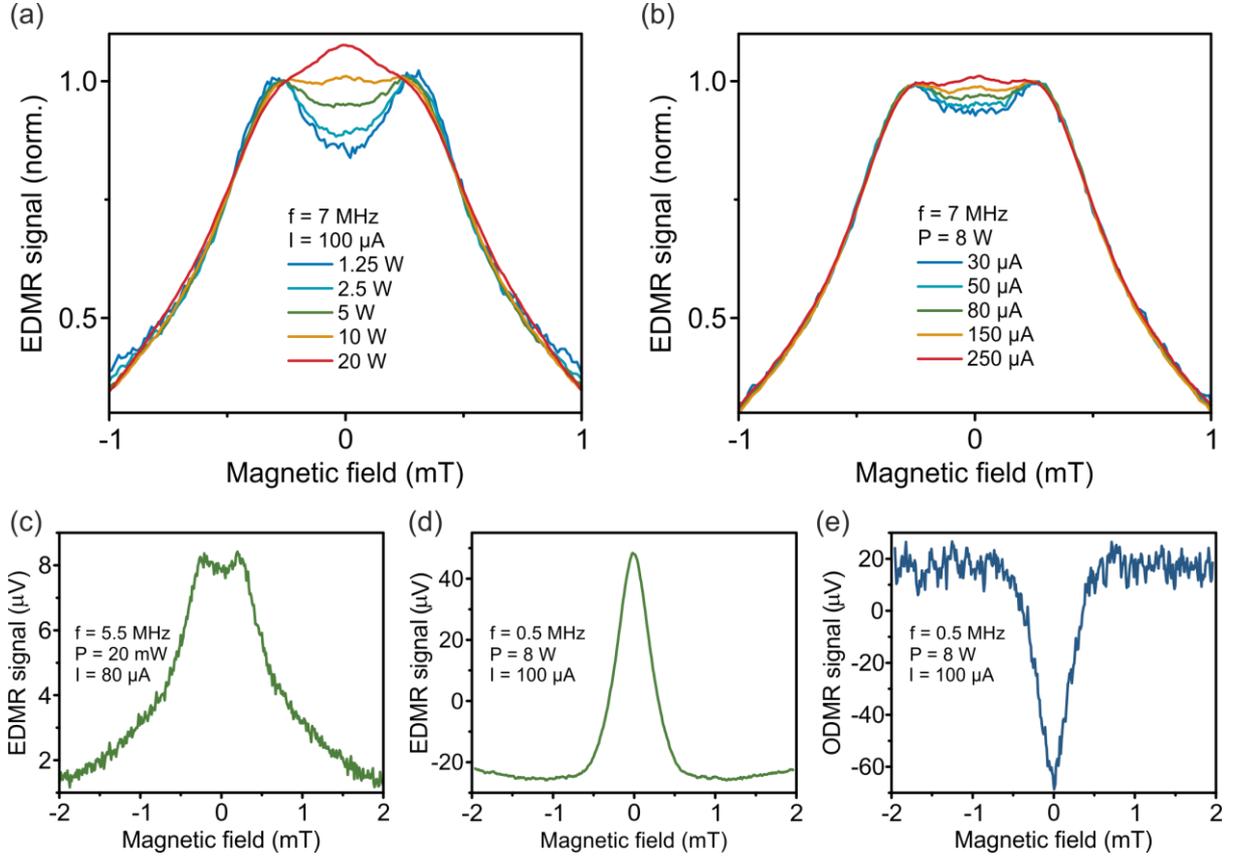

Fig. 8: *Investigation of the zero-field EDMR peak with lock-in detection. (a) Power dependence of EDMR spectra for f = 7 MHz and I = 100 µA. With increasing power, the zero-field peak becomes more pronounced. In (b) similar behaviour is observed for variation of the device current at f = 7 MHz and P = 8 W. Spectra are normalized to the amplitude of the Zeeman resonances in (a) and (b). Optimization of these parameters with respect to the visibility of the Zeeman resonances in (c) allows the detection of g = 2 EDMR as distinct peaks even down to f = 5.5 MHz. Under constant current of I = 100 µA, the zero-field peak is observed without visible contributions from the Zeeman resonances at f = 0.5 MHz and P = 8 W in EDMR (d) and ODMR (e). The width of the zero-field peak matches that of the narrow Gaussian function used to fit the Zeeman resonances at finite fields (see Fig. 5(b)).*

**Magnetic-field effects in a dual-emitting mCP:DMDB-PZ OLED**

The models for avian magnetoception employing the radical-pair picture propose a magnetic-field dependent yield of singlet and triplet reaction products [8, 9]. However, simultaneous experimental access to both products is challenging due to the dark triplet state [11, 48, 49]. To overcome these limitations, we extend our experimental approach to a new class of OLED emitters, which show emission from the first excited triplet state as well as from the singlet. This provides direct access to the magnetic-field dependent yield of singlet and triplet products. As the triplet level is not dipole-coupled to the singlet ground state, perturbation of the spin states is necessary through spin-orbit coupling to enable radiative triplet decay. Conventional heavy-atom based emitters such as iridium complexes showing only triplet emission are not suitable, however, to investigate the magnetic-field induced change of spin-permutation symmetry of charge-carrier pairs. In a previous study, the spin-orbit coupling enhancing effect of heavy atoms was employed by incorporating trace amounts of palladium atoms in the emitter material PhLPPP to detect dual singlet-triplet emission under resonance [27]. While an anticorrelation of singlet and triplet emission was observed for both static as well as dynamic magnetic field effects, it was overshadowed by the emergence of strong keto defect emission over time. Here, we take a different approach by locally enhancing spin-orbit coupling through the choice of appropriate molecular orbitals contributing to the excited state of a metal free emitter molecule. By mixing non-bonding n and $\pi^*$ orbitals, the phosphorescence channel can be opened in OLEDs while retaining fluorescence [50-52].

The material we use is a host-guest system consisting of the emitter molecule 11,12-dimethyldibenzo (*a, c*) phenazine (DMDB-PZ) (Sigma Aldrich) embedded in a 1,3-bis(*N*-carbazolyl)benzene (mCP) (Sigma Aldrich) matrix, the structures of which are shown in Fig. 9a. The two materials are deposited by cosublimation at a ratio of 97:3 of host to guest to yield a 40 nm thick film. A further hole-blocking layer of 20 nm bathophenanthroline was deposited before terminating the device structure with Ba and Al layers as described above. The turn-on voltage of these devices of approximately 7 V is substantially higher than for SyPPV OLEDs. In DMDB-PZ, spin-orbit coupling gives rise to mixing between the molecular orbitals of the lone-pair electrons of the nitrogen atom and the conjugated π-system. This mixes singlet character into the triplet excited state and provides a sufficient transition dipole moment for the radiative transition from the first excited triplet state. In addition to a singlet emission peak around 440 nm, luminescence from this material features a contribution from the triplet excited states, most prominent around 560 nm, as shown in the EL spectrum in Fig. 9b. Because of quenching processes such as triplet-triplet or triplet-polaron annihilation, the spectral weight of triplet emission decreases with increasing device current.

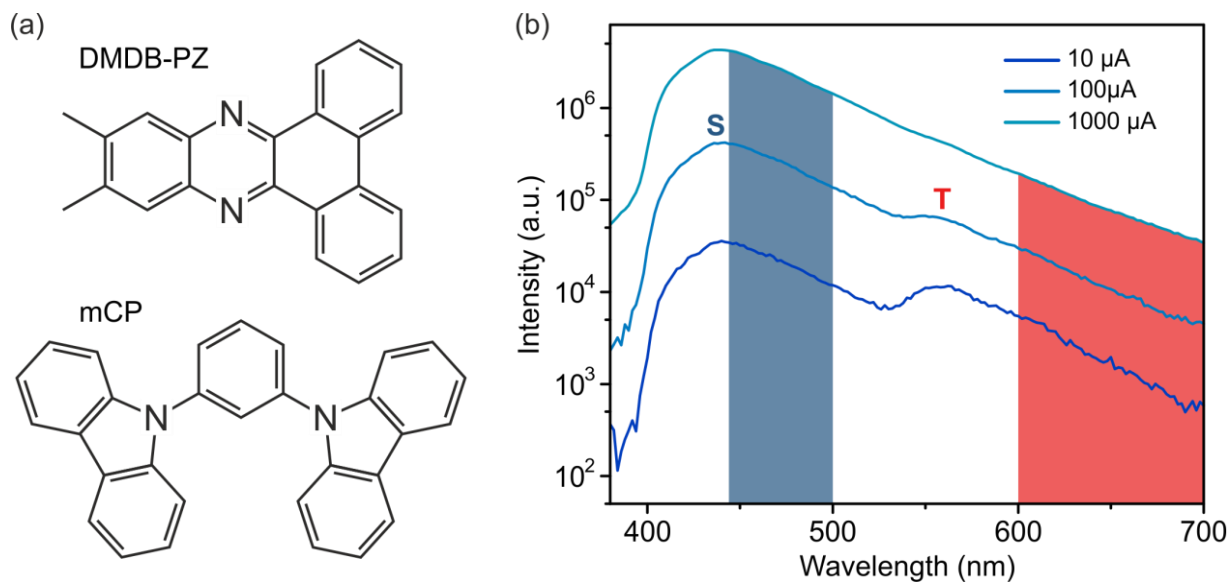

*Fig. 9: (a) Chemical structures of the emitter DMDB-PZ and the mCP matrix. (b) EL spectra of an mCP:DMDB-PZ device at different currents. Singlet emission (S) peaks at around 440 nm, while triplet emission (T) is observed around 560 nm. Due to quenching processes, the triplet emission is less prominent at high current levels. The spectral regions detected with the experimental setup are indicated by blue shading for the singlet channel and red shading for the triplet channel.*

In order to test the predictions of the radical-pair model, the EL intensity in both fluorescence and phosphorescence is recorded as a function of the externally applied magnetic field $B_0$. We spectrally select singlet and triplet emission by splitting the EL with a 552 nm dichroic mirror with a subsequent 500 nm short-pass filter for fluorescence and a 600 nm long-pass filter for phosphorescence. The transmitted light from each channel is incident on an optical power meter (Femto OE-200-SI), which is read out by a Keysight 34461A multimeter. The device is operated under constant-current conditions of $I = 500$ µA, corresponding to a current density of 39 mA/cm$^2$ for the mCP:DMDB-PZ OLED. Upon sweeping the external magnetic field from -23 to +23 mT, we find a clear anticorrelation between the short-wavelength and the long-wavelength emission channel, as displayed in Fig. 10a. For the device resistance under constant-current drive, a decrease with magnetic field is observed, but the effect is much smaller than that observed for SyPPV OLEDs. The observed anticorrelation indicates that, indeed, the hyperfine-induced spin mixing between singlet-like and triplet-like states is suppressed upon the application of a magnetic field. Note that in the general case, the anticorrelation is not necessarily quantitative because singlet as well as triplet MEL are influenced by transport effects [52]. In addition, the spectral overlap between singlet and triplet emission peaks and the precise spectral positions of the optical filters may impact the relative intensity of singlet and triplet luminescence detected.

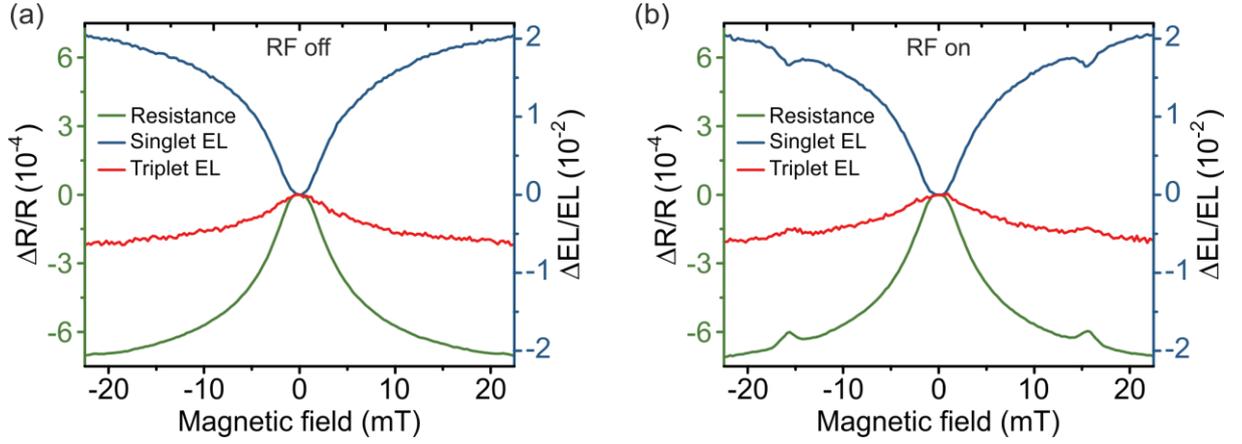

*Fig. 10: (a): Magnetoresistance and MEL of the mCP:DMDB-PZ device. A decrease of device resistance (green) is observed with increasing magnetic field. Singlet (blue) and triplet MEL (red) show the anticipated anticorrelation. Upon RF excitation at $f = 440$ MHz and $P = 20$ W, resonant peaks emerge at $B_{EPR} = 15.7$ mT (b). The anticorrelation between singlet and triplet intensities is retained in ODMR.*

We can directly resolve the singlet-triplet spin mixing under magnetic resonance in OLEDs of mCP:DMDB-PZ with the setup detailed in the discussion of the SyPPV OLEDs. As shown in Fig. 10b, again, resonant peaks are observed in MR as well as in both singlet and triplet MEL under RF excitation with $f = 440$ MHz and $P = 20$ W. To improve the signal-to-noise ratio we resort to lock-in detection with 100% modulation of the excitation amplitude, shown in Fig. 11a. In this case we choose a lower modulation frequency of $f_{\text{mod}} = 23$ Hz (compared to 232 Hz used above) to be able to observe the long-lived triplet emission with the lock-in technique. For both methods of detection − steady-state and lock-in − an anticorrelation between singlet and triplet ODMR is observed, showing that paramagnetic resonance constitutes a channel for mixing between the singlet and the triplet manifold. Again, the anticorrelation of singlet and triplet ODMR peaks is not quantitative, as triplet-triplet and triplet-polaron annihilation induce an effective asymmetry of the mixing, thereby influencing the overall exciton yield.

In contrast to ODMR in the aforementioned material PhLPPP [27], where the formation of keto-type oxidative emissive defects hinders the phosphorescence detection, we find the singlet-triplet anticorrelation to be stable over many hours. This stability allows us to reliably perform an analysis of the lineshape of the EDMR and ODMR spectra. As the influence of spin-orbit coupling on the EDMR and ODMR lineshape can only be obtained from density-functional calculations or high-field magnetic resonance spectroscopy [41, 53], we again resort to the simple model of the sum of two Gaussian functions to describe the inhomogeneous broadening arising from the hyperfine field distributions experienced by electron and hole. Even with this simple model, the experimental data are well approximated, as shown in Fig. 11b. To analyse the resonance linewidths, a total of 28 EDMR spectra as well as 27 singlet ODMR and 18 triplet ODMR spectra are taken into account. For all three channels, good agreement with the resonance field expected for the free-electron g-factor is found. The Gaussian linewidths of the resonance $\Delta B_1$ and $\Delta B_2$, arising from the two carrier species, for the resistance as well as for the singlet and triplet channels are significantly larger than the values obtained for SyPPV, as summarized in Table 1.

There are several possible reasons for this increase. First, the charge carrier wavefunctions are presumably more strongly localized in mCP and DMDB-PZ than in SyPPV, giving rise to stronger overall hyperfine interactions – the more delocalized the carrier spin, the more the average over the isotropic hyperfine fields tends to zero. Second, because of the $n\pi^*$ orbital mixing, spin-orbit coupling may be stronger in DMDB-PZ, which would give rise to an overall greater linewidth. To assess these two effects we need to consider deuterated analogues of mCP:DMDB-PZ and investigate closely the resonance spectra at different magnetic field strengths and resonance frequencies. We note, however, that the effective width of the MR and MEL characteristics, which are both governed by the local hyperfine field distributions [38, 54], are rather similar for SyPPV and mCP:DMDB-PZ, suggesting that the level of hyperfine interactions could be comparable in the two materials. It therefore seems plausible that increased spin-orbit coupling in the dual emitter material is the reason for the broader EDMR and ODMR spectra.

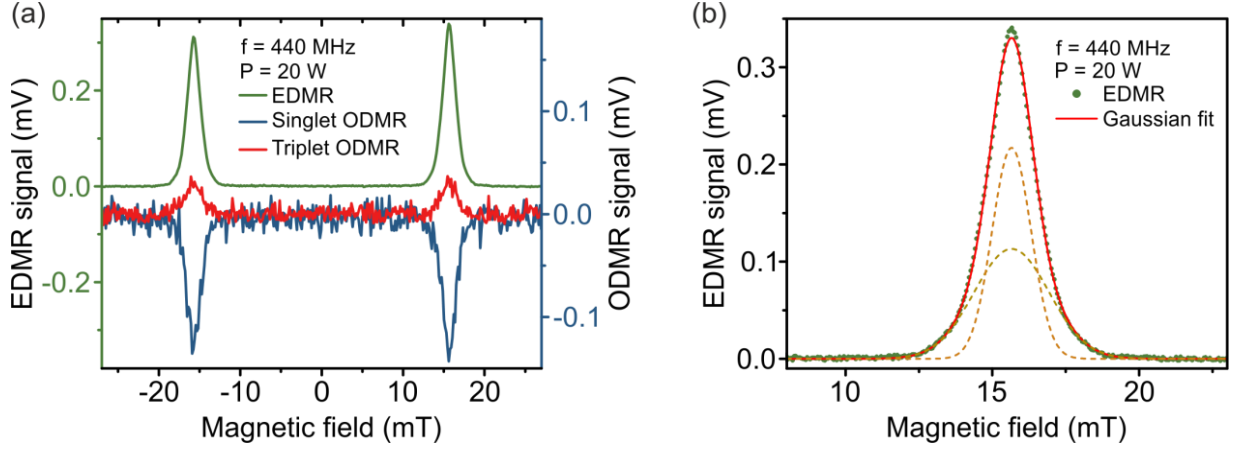

*Fig. 11: Lock-in detection of ODMR and EDMR spectra of mCP:DMDB-PZ. (a) Measurement at f = 440 MHz and P = 20 W. EDMR (green), singlet ODMR (blue) and triplet ODMR (red) with resonance field $B_{EPR}$ = 15.7 mT are detected simultaneously. The resonant signals in the singlet and triplet channel are anticorrelated. (b) Fit of an EDMR curve (green) with the sum of two Gaussian functions (red). The individual Gaussian curves are indicated as dashed lines. For increased visibility, a constant offset is subtracted from the data.*

*Table 1: Comparison of averaged EDMR and ODMR linewidths (FWHM) for SyPPV and mCP:DMDB-PZ OLEDs.*

|  | $\Delta B_1$ (mT) | $\Delta B_2$ (mT) |
| --- | --- | --- |
| SyPPV | 2.31(5) | 0.44(1) |
| mCP:DMDB-PZ | 3.2(4) | 1.6(2) |

Given the possibility of reliable lineshape analysis, we anticipate that we can shed light on an as yet unresolved question in future experiments. As the emitting mCP:DMDB-PZ layer is a host-guest system, it is not clear whether the charge-carrier pairs undergoing magnetic resonance are located on either the emitter or the matrix molecules or whether one carrier sits on the emitter and the other one on the matrix molecule. By using different combinations of hydrogenated as well as deuterated emitter molecules and matrix materials, the distribution of hyperfine fields can be varied [38, 54, 55], leading to a modification of the resonance linewidth. We therefore expect that by careful analysis of EDMR and ODMR spectra it will become possible to pinpoint the location of the individual charge carriers on either emitter or matrix molecules.

## Conclusions

In summary, we have shown that magnetic-field sensitivity of an OLED to changes in the magnetic field strength on the nanotesla scale can be achieved at room temperature. The spin-coherence times necessary to provide such sensitivity at room temperature approach the millisecond regime, which is similar to the timescales invoked to explain avian magnetoception. The underlying radical-pair mechanism responsible for this magnetic-field sensitivity was tested by magnetic resonance measurements with Zeeman resonances being resolved down to the frequency scale of a few MHz. At these frequencies, an additional peak stemming from resonances arising in the isotropic hyperfine fields is observed. Further confirmation of the radical-pair model in the context of OLEDs is obtained by the observation of an anticorrelation of ODMR spectra detected in fluorescence and phosphorescence in a metal-free dual emitter OLED system, corroborating the picture of the transfer of charge-carrier pairs of well-defined spin multiplicity into molecular excitations in either the singlet or the triplet manifold.

Finally, we are confident that, with further optimization of materials to reduce the overall hyperfine field strengths as well as of devices to minimize the RF power and OLED current necessary to detect EDMR, it will be possible to observe features of magnetic resonances directly in OLEDs exposed to Earth's magnetic field: at present, the lowest resonance frequency that was resolved clearly is 5.5 MHz, whereas resonances due to the geomagnetic field are expected to arise around 1.3 MHz. Detecting magnetic resonance in an OLED in geomagnetic fields would provide an appealing complementary alternative to behavioural studies of birds [10, 13, 14] in probing radical-pair mediated processes. Employing dual-emitting materials might offer the unique possibility to directly assess the effect of magnetic fields on singlet and triplet product yield.

## Conflicts of interest

There are no conflicts of interest to declare.

## Acknowledgements

22The authors would like to thank Dr. Vagharsh Mkhitaryan and Dr. Hans Malissa for insightful discussions as well as Sebastian Krug for technical support. We gratefully acknowledge funding from the DFG through the SFB 1277 project B03.